\documentclass[useAMS,usenatbib]{mn2e}
\usepackage{txfonts}
\usepackage{graphicx}
\newcommand{\coeighteentwo}{${\rm C^{18}O}~J=2-1$}
\newcommand{\hcofour}{${\rm HCO^+}~J=4-3$}
\newcommand{\hcothree}{${\rm HCO^+}~J=3-2$}
\newcommand{\hcoone}{${\rm HCO^+}~J=1-0$}
\newcommand{\cstwo}{${\rm CS}~J=2-1$}

\begin{document}
\title[Oscillations in B68]{Oscillations in the stable starless core Barnard 68}
\author[M.P. Redman et al]{M.P. Redman$^{1,2,3}$, E. Keto$^{4}$, J.M.C.
Rawlings$^{3}$\\ 
$^1$Department of Physics, National University of Ireland
Galway, Galway, Ireland.\\ 
$^2$School of Cosmic Physics, Dublin Institute for Advanced Studies, 5 Merrion
Square, Dublin 2, Ireland.\\ 
$^3$Department of Physics \& Astronomy, University College London, Gower
Street, London WC1E 6BT UK.\\ 
$^4$Harvard-Smithsonian Center for Astrophysics, 60 Garden Street, Cambridge 
MA 02138, USA.}

\date{\today}
\pubyear{2006} 
\volume{000}
\pagerange{\pageref{firstpage}--\pageref{lastpage}}
\maketitle 
\label{firstpage}

\begin{abstract}
New molecular line observations of the Bok globule Barnard 68 in HCO$^+$ 
irrefutably confirm the complex pattern of red and 
blue asymmetric line profiles seen across the face of the cloud in previous 
observations of CS.  The new observations thus strengthen
the previous interpretation that Barnard 68 is undergoing
peculiar oscillations.
Furthermore, the physical chemistry of B68 indicates that the object is much 
older than the sound crossing time and is therefore long-lived. A model is 
presented for the globule in which a modest external pressure perturbation is 
shown to lead to oscillations about a stable equilibrium configuration.  Such 
oscillations may be present in
other stable starless cores as manifested  by a similar signature of
inward and outward motions. 
\end{abstract}


\section{Introduction}
One difficulty in identifying a potential site of future star
formation has been to decide whether a dense portion of a molecular
cloud is likely to proceed to collapse to form a star or else is
just a transient or a long-lived density enhancement. Models of supersonically
turbulent cold molecular clouds readily produce localised density
enhancements in colliding gas streams
(e.g. \citealt{klessen.et.al00,padoan&norlund02,ballesteros-paredes.et.al03}). 
However, in the simulations, the clouds  either quickly dissipate on a sound
crossing time or, if their density and size are such that the cloud mass 
exceeds the local Jeans mass, they may
collapse to form  stars.  However, observations that show that the internal 
structure of some clouds is consistent with
self-gravitating equilbrium suggest that the alternative of long-lived clouds
should not be excluded. Unbiased observations of large swathes of molecular 
cloud gas point to cloud survival timescales a few sound crossing times but 
less than order ten \citep{visser.et.al02}.
Theoretical models of small clouds in equilibrium closely match the
observations and suggest that small clouds may be catagorized as either `stable
starless cores' or `unstable pre-stellar cores' using the terminology suggested in Keto \& Field (2005)  (note that the terms `starless cores', `pre-protostellar' and `pre-collapse' cores are commonly used in observational work). 
Such structures, in which local thermal pressure dominates over larger scale
inertial forces, could represent the end point of the turbulent cascade in
larger scale molecular clouds.
Thus the picture of the interstellar medium as a purely turbulent phenomenon,
and all clouds as purely transient  may not be applicable on the scale of the
smaller clouds
such as Bok globules and starless cores (Fig 1). Since star formation in regions such
as Taurus and Ophiuchus takes place in these small clouds, 
an understanding of whether equilibrium conditions can prevail is crucial in 
order to be able to understand the initial conditions of star formation and the
firm identification of infall candidates. 

Barnard 68 (LDN 57, CB 82 hereafter B68, Figs ~{\ref{b72} and \ref{b68grey})
is an excellent test case of an object that could either be a stable
starless core or unstable pre-stellar core. Because B68 is nearby 
($\sim125~{\rm pc}$) and fortuitously seen in silouette against the galactic 
bulge, Alves et al (2001a) were able to directly measure
the dust extinction in the core with unprecedented accuracy using deep near-infrared extinction measurements from H and K imaging of 
background stars.
With the assumption of a uniform gas to dust ratio, the gas density profile 
for this core is thus very well constrained as a cloud in hydrostatic
equilibrium confined by external pressure: a Bonner-Ebert Sphere
\citep{bonner56,ebert55} which is a solution of the 
Lane-Emden equation for a bounded isothermal pressure-balanced sphere. However, we 
argue below that B68 is not isothermal and is possibly not is hydroststatic
pressure balance.
The high quality of the absorption data and the precise determination of 
the column density removes a major source
of uncertainty for investigating the internal structure of this
possible prestellar core candidate.

\begin{figure} 
\includegraphics[width=225pt,bb=20 20 575 365]{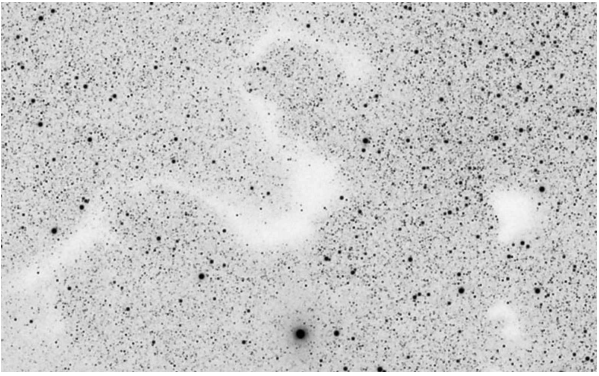}
\caption{Barnard 68 (lower right) and nearby molecular clouds including Barnard 
72, the snake nebula. Image credit and copyright Gary Stevens}
\label{b72}
\end{figure} 
\begin{figure} 
\includegraphics[width=225pt,bb=20 20 575 555]{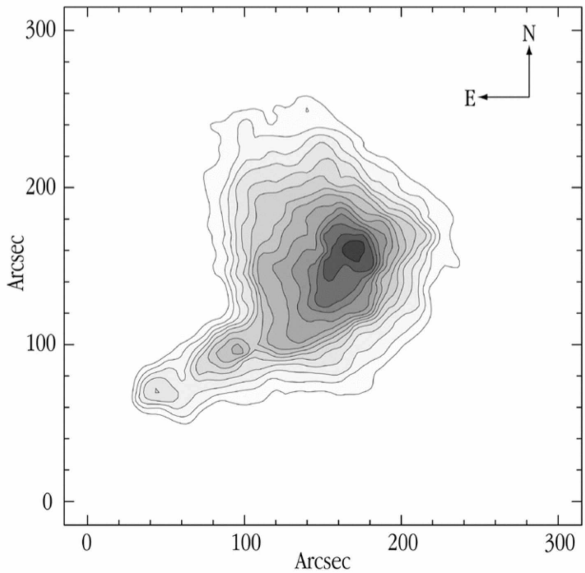}
\caption{Iso-extinction map of B68 from Alves et al. (2001b). Image credit 
and copyright ESO}
\label{b68grey}
\end{figure} 
The dust obscuration map in Fig~\ref{b68grey} \citet{alves.et.al01b} show that B68 consists of an approximately
spherical component of radius 12500~{\sc au} with a markedly off-centre dust 
peak. There is also a lower density extension to
the south-east.  \citet{alves.et.al01a} estimated that the central
density of B68 to be $\sim 2.5\times 10^5~{\rm cm^{-3}}$ giving a total mass of 
the cloud of $\sim 2 M_\odot$. Note that \citet{hotzel.et.al02a} argue that B68 
is substantially closer ($\sim$80pc)
than the 125 pc used by \citet{alves.et.al01a} which would make the cloud 
smaller and less massive.

The \citet{alves.et.al01a} paper has led to intense interest in B68 and
its basic physical and chemical properties have now become well constrained.
Several measurement of the gas and dust temperatures have been made. ISOPHOT 
observations by \citet{langer&willacy01}
observations indicate a cold central dust temperature of 7K surrounded
by a warmer envelope. \citet{hotzel.et.al02b} use ammonia inversion
lines to measure a gas kinetic temperature of $10\pm 1.2~{\rm
K}$. Similarly, \citet{lada.et.al03} find a gas temperature of
$10.5~{\rm K}$ from $^{12}$CO measurements in B68. These results are entirely consistent with continuum submillimetre  observations and radiative transfer models of this and other sources of  similar evolutionary age \citep{evansetal01}. These models clearly show that the cores possess significant temperature  structure, with a warmer ($\sim$10-15~K) envelope surrounding a colder ($\sim$5-8~K) inner region.

The above temperatures are also consistent with molecular observations:
The species used for calculating the gas temperature show evidence for heavy 
depletion due to freeze-out onto dust grains in the central regions of B68. 
Indeed, dust emissivity measurements by \citet{bianchi.et.al03} point to ice 
covered coagulated grains in the core of B68. The depleted species include 
${\rm H_2O}$, ${\rm NH_3}$, ${\rm CO}$, CS and even ${\rm N_2H^+}$
(\citealt{bergin.et.al02, 
bergin&snell02,hotzel.et.al02a,difrancesco.et.al02,lai.et.al03}). For reasons that are still not clear, N$_2$H$^+$ is found to be remarkably
resistant to freeze-out and traces matter to surprisingly high densities in many
cloud cores. The chemistry of N$_2$H$^+$ is relatively simple; it is formed by reaction of N$_2$ with H$_3^+$ and destroyed by dissociative recombination and reaction with 
CO. The main loss route for CO itself being reaction with H$_3^+$. The high
N$_2$H$^+$ abundances that are seen in many regions of modertae to high 
depletion has previously been ascribed to the low surface binding energies of
N$_2$ and related species, but recent laboratory and theroretical work suggests
that the binding energies are probably larger than previously thought. In any 
case, at very high densities and after long periods of time,
even N$_2$H$^+$ will freeze-out. In B68 \citet{bergin.et.al02} estimate that 
N$_2$H$^+$ is depleted by a factor of 2 (between A$_v\sim 2-17$), whilst 
\citet{difrancesco.et.al02} suggest that the level of depletion may be an order
of magnitude larger. The central density of B68 is not anomalously 
high, so the obvious conclusion is that the core has been quasi-static for an 
exceptionally long period of time.

The chemical determination that  B68 has a long lifetime is important in understanding the evolution of the cloud because both stable and unstable Bonnor-Ebert spheres have an approximate balance of thermal and gravitation forces, provided the clouds are not in free fall \citep{keto&field05}. Therefore, observations of structure consistent with an approximate balance of forces is an ambiguous indicator of the evolutionary fate of a cloud.

The observations of starless cores such as B68 often show spectral line
profiles indicating large scale gas motions. If interpreted in terms of simple
contraction or expansion, the clouds could not have lifetimes longer than a
crossing time. If a starless core can exhibit significant internal motions yet
not ultimately collapse or evanesce, this would have profound implications
for the continuing attempts to firmly identify infalling prestellar
cores and test the competing dynamical models of star formation.

\begin{figure*} 
\includegraphics[width=450pt,bb=20 20 575 142]{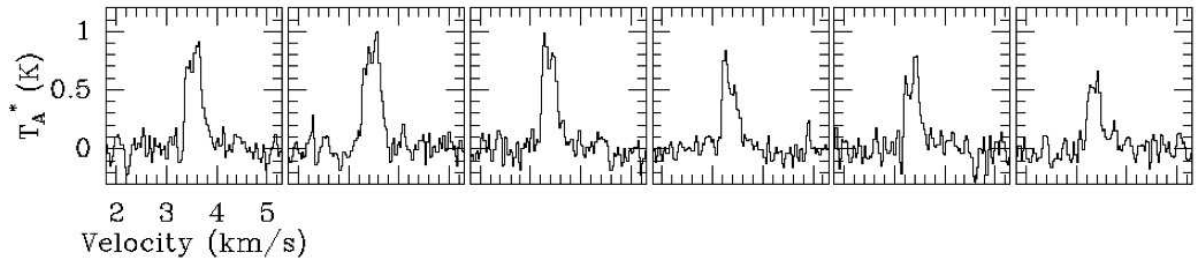}
\caption[ladaprofiles]{Lada et al (2003) line profiles of \cstwo\ across
a strip through B68. The profiles are separated by 24 arcseconds.}
\label{ladaprofiles}
\end{figure*}

The spectral line observations of 
\citet{lada.et.al03} 
show profiles that are all double peaked and asymmetric but, most remarkably,
they alternate between being red asymmetric and blue asymmetric across
the core. These observations
consist of  \cstwo\  line profiles and a velocity centroid map of 
\coeighteentwo.  
In this paper we present additional molecular line observations of \hcothree\ 
in a two-dimensional pattern covering the face of B68. In sections~\ref
{observations} these profiles are presented and compared with the data of Lada 
et al (2002). Section~\ref{discussion} describes a dynamical model for this 
cloud to account for the shapes of the line profiles and conclusions are drawn 
in section~\ref{conclusions}.

\section{Observations} 

\label{observations} JCMT observations of the \hcothree\ line were
collected gradually in service observing mode over the year August
2003 to August 2004. In addition to observing across the east-west
strip of points examined by \citet{lada.et.al03}, data were also
gathered from several other strips to build up a map of 38 separate
pointings. The \citet{lada.et.al03} \cstwo\ data are reproduced in 
figure~\ref{ladaprofiles}. Figure~\ref{pointings} shows the location of the 
JCMT pointing positions against the \citet{alves.et.al01a} image of B68. 
Good signal to noise at each point was preferred over
additional pointings in order that the substructure of the line
profile could be adequetely detected.  The data were reduced in the
standard manner using the {\sc specx} package in the {\sc starlink}
suite of astronomical software. A beam efficiency of 0.65 was used to convert the temperatures to the $T_{\rm mb}$ scale.
\begin{figure} 
\includegraphics[width=210pt,bb=20 20 575 570]{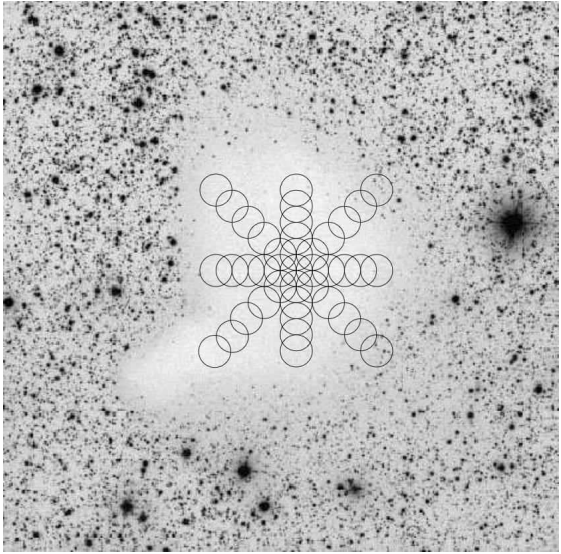}
\caption{Pointing positions of JCMT shown projected against reverse greyscale 
ESO optical image of B68 (Alves et al 2001). The circles represent the 15" 
beamsize of the JCMT.}
\label{pointings}
\end{figure}

Figure~\ref{map} is a map of the \hcothree\ line profiles. Individual
panels are separated by 10 arcseconds to the north and/or east. The central 
east-west profiles
are at the same declination as the \citet{lada.et.al03} profiles
reproduced in Fig~\ref{ladaprofiles}. Comparing the profiles it is
immediately obvious that the pattern of asymmetry changes in the
\cstwo\ line observed by \citet{lada.et.al03} is present in exactly
the same way in the \hcothree\ line. The rest of the data show that the 
distribution of the asymmetric profiles exhibits some overall order in that 
there are large contiguous blocks of red or blue asymmetric profiles.
\citet{lada.et.al03} also used the difference of the centroid
velocities between their ${\rm C^{18}O}$ and \cstwo\ data to construct
a map of the asymmetries across the face of B68. Again, the line
profiles obtained here match this distribution of red and blue
asymmetries closely. The fact that species/transitions with different chemical behaviours and excitation characteristics apparently trace the same kinematics is indicative of bulk motions propagating deep into the core (but traced no deeper than the freeze-out radius: cf the ${\rm N_2H^+}$ data of Lada et al 2003, which exhibit somewhat different kinematics extending to the very centre of the core), rather than disturbances in the surface layers. In this context, it is interesting to note (see 
Figure 6) that the velocity of the self-absorption dips does not vary 
significantly across the source. This would again suggest that the origin of
the kinematics and the reversal of the line profile asymmetry is due to dynamical activity extending deep into the cloud. 
\begin{figure*} 
\includegraphics[width=400pt,bb=20 20 575 545]{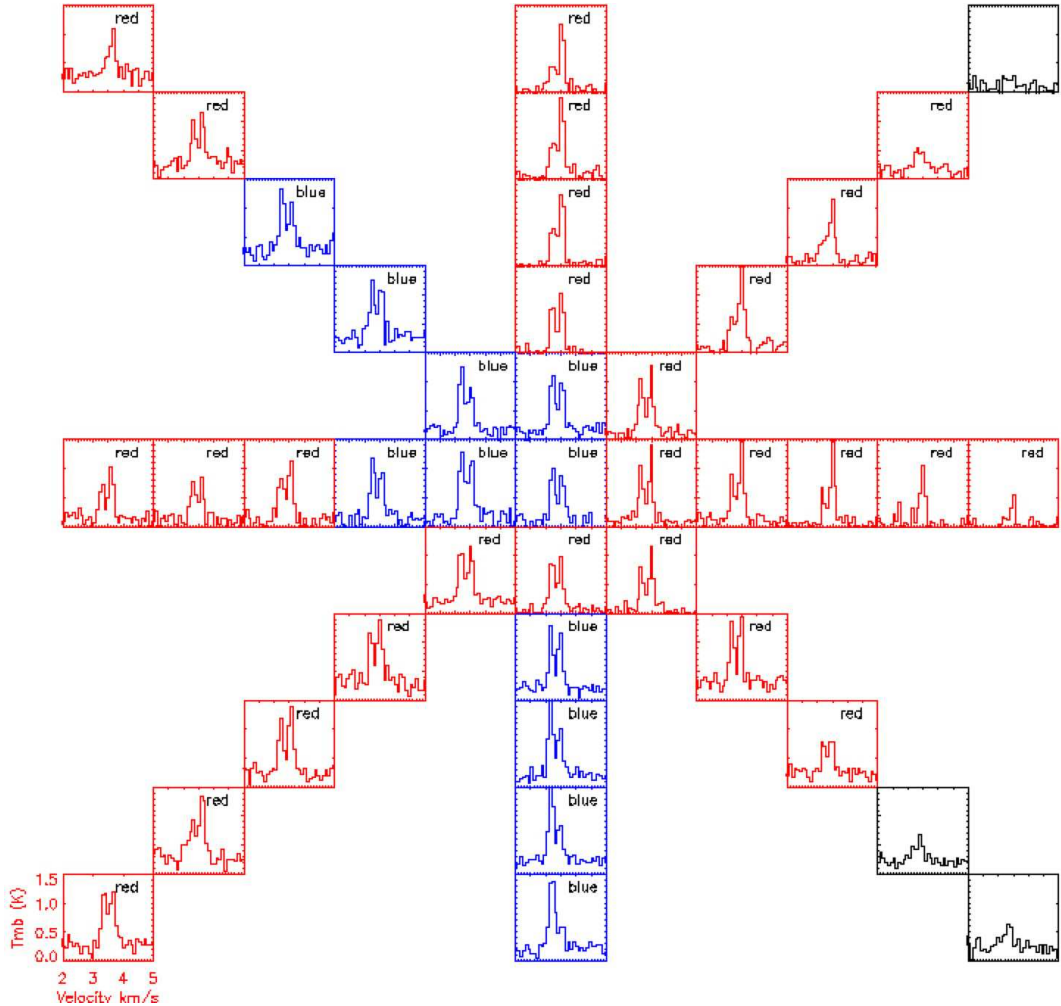}
\caption{JCMT line profiles of \hcothree\ across B68.}
\label{map}
\end{figure*}
\begin{figure} 
\includegraphics[width=200pt,bb=20 20 575 548]{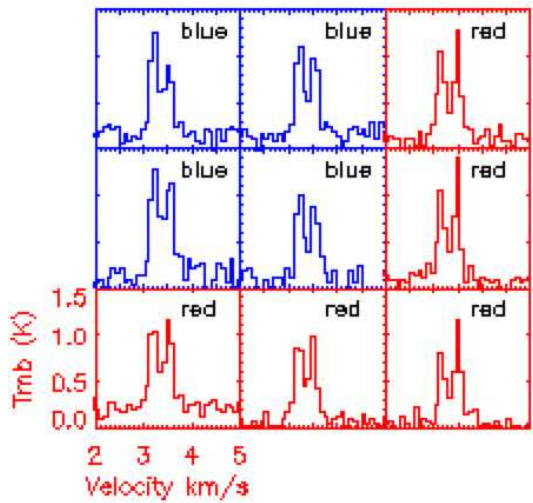}
\caption{The central nine profiles of the JCMT data from the previous
figure allowing detail of the asymmetry to be seen more clearly}
\label{map2}
\end{figure}

\section{Discussion and model}
\label{discussion}

The observed velocity pattern is very hard to explain with simple combinations 
of infall, outflow or rotation (or depletion). For an optically thick cloud 
undergoing collapse, blue asymmetric line profiles are expected everywhere. For 
a rotating cloud with no infall, an ordered set of red asymmetric and blue 
asymmetric profiles are obtained either side of the rotation axis 
\citep{redman.et.al04a}. The red and blue asymmetric profiles cannot be caused 
by an outflow because there is no central source to power one present in B68. 
Because the pattern of asymmetry is present in two different tracer species 
then, as described above, it is likely an effect of gas motion i.e. a dynamical effect as opposed to an effect which varies by species such as line generation or depletion. So a new mechanism is required that could produce strong enough dynamical
activity that the cloud can exhibit such readily observable asymmetric
profiles. 

Keto \& Field (2005) analysed the hydrodynamical response of  
Bonner-Ebert spheres to external pressure disturbances. They found that stable starless cores (i.e. those on the stable branch of the Bonner-Ebert sphere solution curve) can undergo bulk compression or expansion motions that do not tip the core into gravitational collapse, provided the amplitude of the perturbation is not too large. Such a model naturally explains the observed origin of the pulsation as deriving from motions within the inner parts of the core.

We model B68 as a Bonnor-Ebert sphere internally supported by thermal pressure and initially in 
pressure equilibrium with the external medium.  The cloud is initially in critical equilibrium and stable against  
gravitational collapse.
 The initial conditions of the model cloud
 are listed in  Table~\ref{parameters}. 
 We then follow the evolution of the cloud in two different
 scenarios. In the first case the external pressure increases by a factor of 4 
and is held
 constant thereafter, and in the second case the pressure decreases by a factor 
of 4.  The resultant hydrodynamics are treated one dimensionally for 
simplicity,  and the density, velocity, pressure and temperature are 
calculated. To compare the model with observations, a  
 molecular line radiative transfer code  \citep{keto&field05,redman.et.al04a} 
is employed to calculate the spectral line profiles of  HCO$^+$ across the 
cloud assuming constant abundance of HCO$^+$.
 
\begin{table}
\begin{tabular}{ll}
Parameter & Adopted value or range \\
\hline
Outer radius & $0.09~{\rm pc}$ (18500 au)\\
Central density & $2\times 10^5~{\rm cm^{-3}}$\\
$\rm HCO^+$ abundance & $5.0\times 10^{-9}$\\
Turbulent velocity & $0.15~{\rm km~s^{-1}}$ \\
Initial gas temperature range & $8-11~{\rm K}$\\
Initial dust temperature range & $13-16~{\rm K}$\\
\hline
\end{tabular}
\label{parameters}
\caption{Model parameters used. The temperature, density,  and velocity fields
which are determined from the numerical hydrodynamic code 
vary as a function of radius within the ranges listed in the table.}
\end{table}

Sample results of the calculations are presented in
Figure~\ref{modelprofiles}. The results show that the 1D cloud model can 
exhibit profiles that are either red or blue asymmetric depending on whether
contraction or expansion is taking place. The \hcothree\ spectral line profiles 
shape closely match our observed spectra and our model also predicts the 
profiles of the \hcoone\ and \hcofour\ lines. We make the firm prediction that 
observations of Barnard 68 taken in
the \hcoone\ line should reveal an identical asymmetry pattern as
pronounced as that in the \hcothree\ data presented here and that the \hcofour\ 
line will be weak and single peaked. Calculations were also performed for the 
CS molecule and similar results (not shown here) obtain and match the \cstwo\  
data of \citet{lada.et.al03}. 

\begin{figure} 
\includegraphics[width=250pt,bb=20 20 575 142]{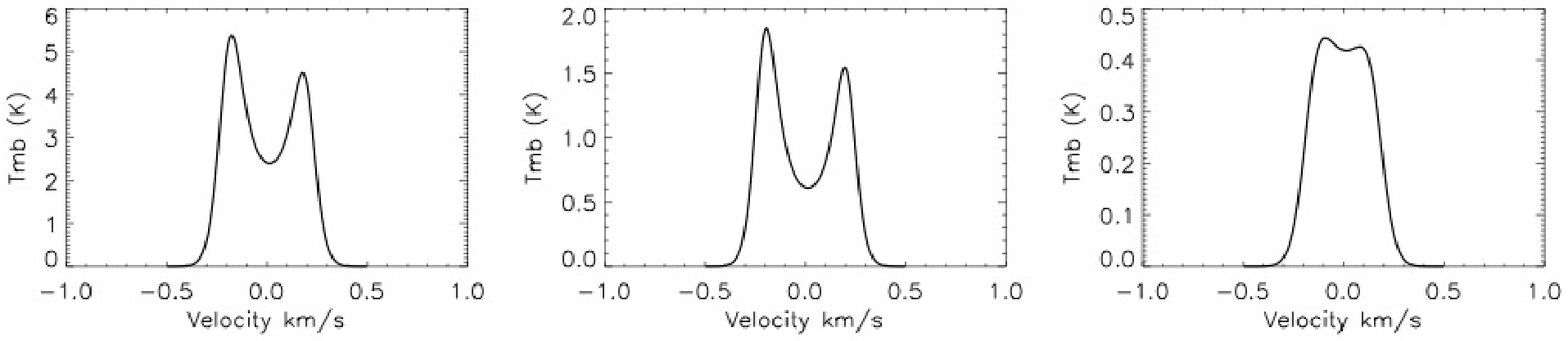}
\includegraphics[width=250pt,bb=20 20 575 146]{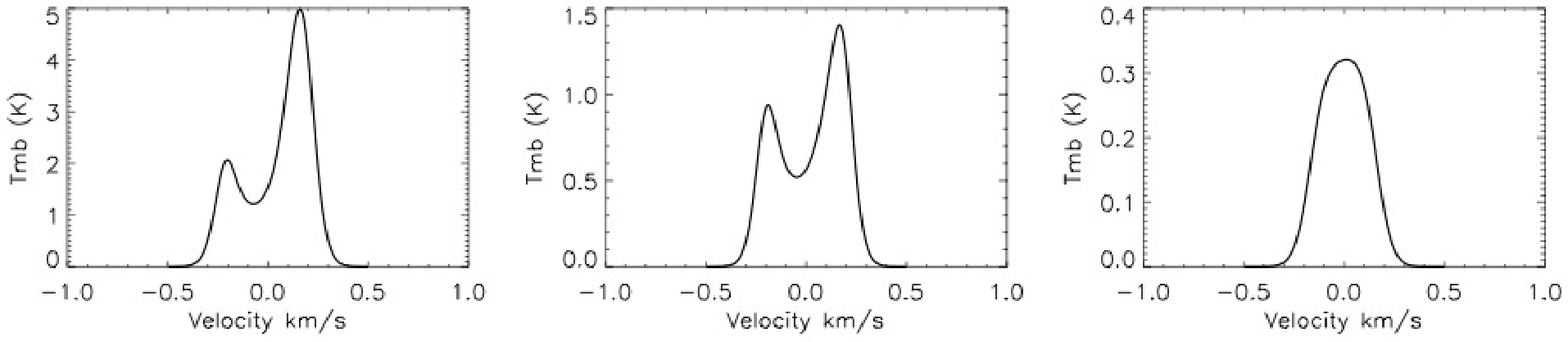}
\caption{Model line profiles of HCO$^+$ for, from left to right, 
$J=1\rightarrow 0$, $J=3\rightarrow 2$ and $J=4\rightarrow 3$. The core is 
stable against overall gravitational collapse yet the pressure increase (upper 
panel) or decrease (lower panel) models exhibit asymmetric line profiles}
\label{modelprofiles}
\end{figure}

The data show a mixture of red and blue asymmetric profiles simultaneously 
across the cloud rather than a single set of expanding or contracting motions
but it seems reasonable that in a real cloud different portions of the
cloud could be in different phases of the cycle. These higher order non-radial 
modes may be excited in the cloud by external pressure perturbations, similar 
to the mechanism envisaged by \citet{lada.et.al03} to explain their results. 
Keto et al (2006, in prep) have calculated the line profiles that would result from an 
analytic model of modes of oscillation and find that zones of expansion and 
contraction produce red and blue assymmetric spectra similar to those seen in 
B68.

It should soon be possible to attempt a fully 3D analysis of B68 in which the 
hydrodynamics, the dust continuum radiative transfer, chemistry and molecular 
line radiative transfer are all carried out self-consistently. Most of the 
ingredients are already available: a 3D hydrodynamic code could be combined 
with the 3D dust density distribution (recently constrained by J. Steinacker et 
al, in prep). Dark cloud chemistry models are available that are begining to 
include freeze-out and the radiative transfer can be calculated in 3D with our 
code.

\section{Conclusions}
\label{conclusions} 

New observational data confirms that B68 exhibits split 
asymmetric molecular line profiles that indicate significant dynamical activity in the 
core. We favour a simple model in which the variety of spectral line profiles 
across the face of the cloud are the result of oscillatory motions about a state
of gravitational equilibrium. The oscillations could have been started by a 
disturbance in the external pressure. A simple numerical hydrodynamic model in 1D that follows the evolution of an equilibrium Bonnor-Ebert sphere
subject to a change in external pressure reproduces
spectral line profiles similar to those observed. This simple model suggests
that a more complex model that allowed for 3D oscillations could reproduce the variety of spectral line profiles across the face of the cloud. 

At a very general level, this study re-inforces the need for caution,
as emphasised by \citet{rawlingsandyates} and others, in the interpretation of
line profiles observed along single lines of sight. A variety of dynamical
activities, including rotation, non-spherical outflows and - as shown in this
paper - pulstation, can mimic the characteristics of simple spherically
symmetric inflow.

The depletion of molecular species observed toward the center of the cloud suggests that the cloud must have a lifetime long enough, a few million yr, to allow collisional processes to deplete the molecules from
the gas phase by freeze-out onto grains. Turbulent models for the interstellar medium tend to indicate rapid star formation as localised
dissipation of energy leads to stars `condensing' out of molecular
clouds. Typically in these models the star formation process is not
traced beyond the formation of a Jeans mass worth of cloud because of
resolution issues; the star
is assumed to appear on a collapse timescale once a suitable
prestellar core has formed. This process of density change is not
neccesarily one way. In another picture, slow mode MHD waves allow for
the formation and then dissipation of density enhancements
\cite{fandh02,garrodetal05}. The chemistry of some
molecular clouds appears to support this latter view in that the
presence of several species is most plausibly explained if the cloud
has undergone several such enhancements and contractions.

The combination of the oscillatory motions, an internal structure consistent
with equilibrium, and a lifetime greater than several crossing time suggests 
that long lived pressure supported clouds may
exist in the turbulent ISM. Some part of the inefficiency of star formation may 
be due to formation of long-lived clouds. Many objects like B68 will 
promptly condense out out a turbulent medium, marking the endpoint of the 
supersonic turbluence cascade, but only a subset are destined to continue to 
collapse all the way to becoming a star. The rest will remain as stable 
starless cores eventually to be dissipated or triggered into collapse by 
external effects due to nearby massive stars and supernovae. The observational challenge is to identify those objects
most likely to collapse with the difficulty, shown here, that some
stable starless cores may exhibit strong evidence of dynamical
activity yet, paradoxically, be perfectly stable against collapse.

\section*{Acknowledgements} 
We thank the referee, C. Lada for a helpful report. MPR was supported by the UK PPARC, and Ireland IRCSET during the early
stages of this work. We thank the staff of the JCMT and visiting
observers for obtaining the observations. The JCMT is operated by the
JAC, Hawaii, on behalf of the UK PPARC, the Netherlands NWO, and the
Canadian NRC. We thank Gary Stevens and Charlie Lada for permission to reproduces figures 1 and 3.

\label{lastpage}
\end{document}